\input{aipcheck}

\documentclass[
    ,final            
  ]
  {aipproc}

\usepackage{bm}

\layoutstyle{8x11single}

\begin{document}

\title{A New Measurement of $\eta_b(1S)$ From $\Upsilon(3S)$ Radiative Decay at CLEO}


\author{Sean Dobbs (for the CLEO Collaboration)}{
  address={Northwestern University, Evanston, IL 60208, USA}
}

\begin{abstract}
Using CLEO data, we report on the confirmation of the $\eta_b(^1S_0)$ ground state of bottomonium in the radiative decay $\Upsilon(3S)\to\gamma\eta_b$.  We determine its mass to be $M(\eta_b) = 9391.8 \pm 6.6 \pm 2.1$~MeV, which corresponds to the hyperfine splitting $\Delta M_{hf}(1S)=68.5\pm6.6\pm2.0$~MeV, and the branching fraction $\mathcal{B}(\Upsilon(3S)\to\gamma\eta_b)=(7.1\pm1.8\pm1.1)\times10^{-4}$.  These results agree with those previously reported by BaBar.
\end{abstract}

\maketitle

\section{Introduction}

Bottomonium ($\left|b\bar{b}\right>$) is the preferred system for the study of the $q\bar{q}$ interaction.  Theoretical calculations for bottomonium are generally more reliable than for other $q\bar{q}$ systems due to the small relativistic corrections ($v/c\approx0.1$).
However, much about the spectrum of bottomonium states remains unknown.  Although potential models predict $\sim26$ bottomonium states to lie below the $B\overline{B}$ threshold (as illustrated in Fig.~1), only 10 of them had been identified since the discovery of bottomonium in 1977~\cite{ups-disc} until 2008.  Most importantly, the ground state $\eta_b(1^1S_0)$, had not been identified, despite many searches by CUSB~\cite{etabcusb} and CLEO~\cite{ups-cleo} at CESR, and ALPEH~\cite{etabaleph} and DELPHI~\cite{etabdelphi} at LEP, among others.  

This situation changed in July 2008, when the BaBar Collaboration announced the discovery of $\eta_b(1S)$ in their data for 109~million~$\Upsilon(3S)$~\cite{ups-babar1,ups-babar2} in the radiative decay $\Upsilon(3S)\to\gamma\eta_b(1S)$, with a significance of $>10\sigma$, as illustrated in Fig.~2.  We note that the small peak at $E_\gamma\approx920$~MeV due to the radiative transition to $\eta_b(1S)$ is not visible until the large smooth background is subtracted.  BaBar's success is due not simply due to their large data set, more than an order of magnitude larger than previously collected at CLEO, but also due a factor 3 greater background suppression by the use of a cut on the ``thrust angle''.  The thrust angle is defined as the angle between the signal transition photon candidate and the thrust axis of the rest of the event~\cite{thrust}.

BaBar's discovery of $\eta_b$ is exciting and like all such claims, needs to be confirmed by an independent experiment.  We report on such a measurement at CLEO.

\begin{figure}
\includegraphics[width=2.8in]{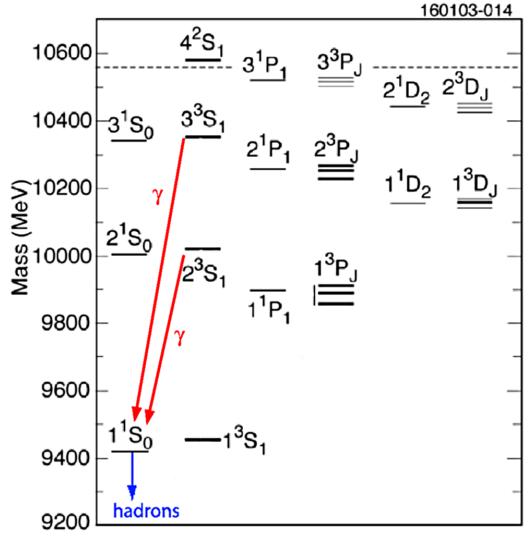}
\caption{Spectrum of the bound bottomonium ($b\bar{b}$) states.  The M1 transitions from the $\Upsilon(n^3S_1)$ states to the $\eta_b(1^1S_0)$ are indicated by red arrows for the transitions reported in this paper.}
\end{figure}

\begin{figure}
\includegraphics[width=2.4in]{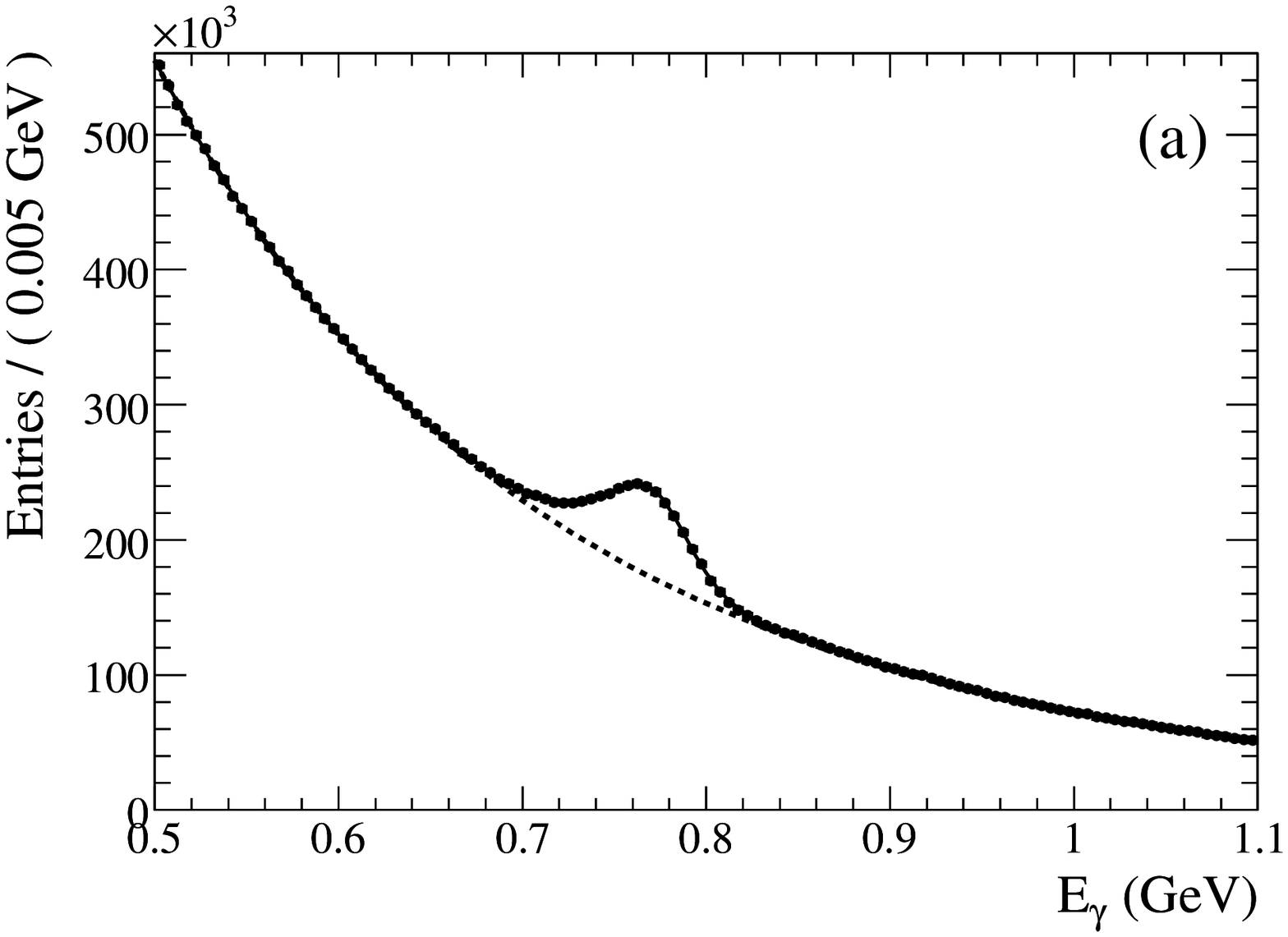}
\includegraphics[width=2.5in]{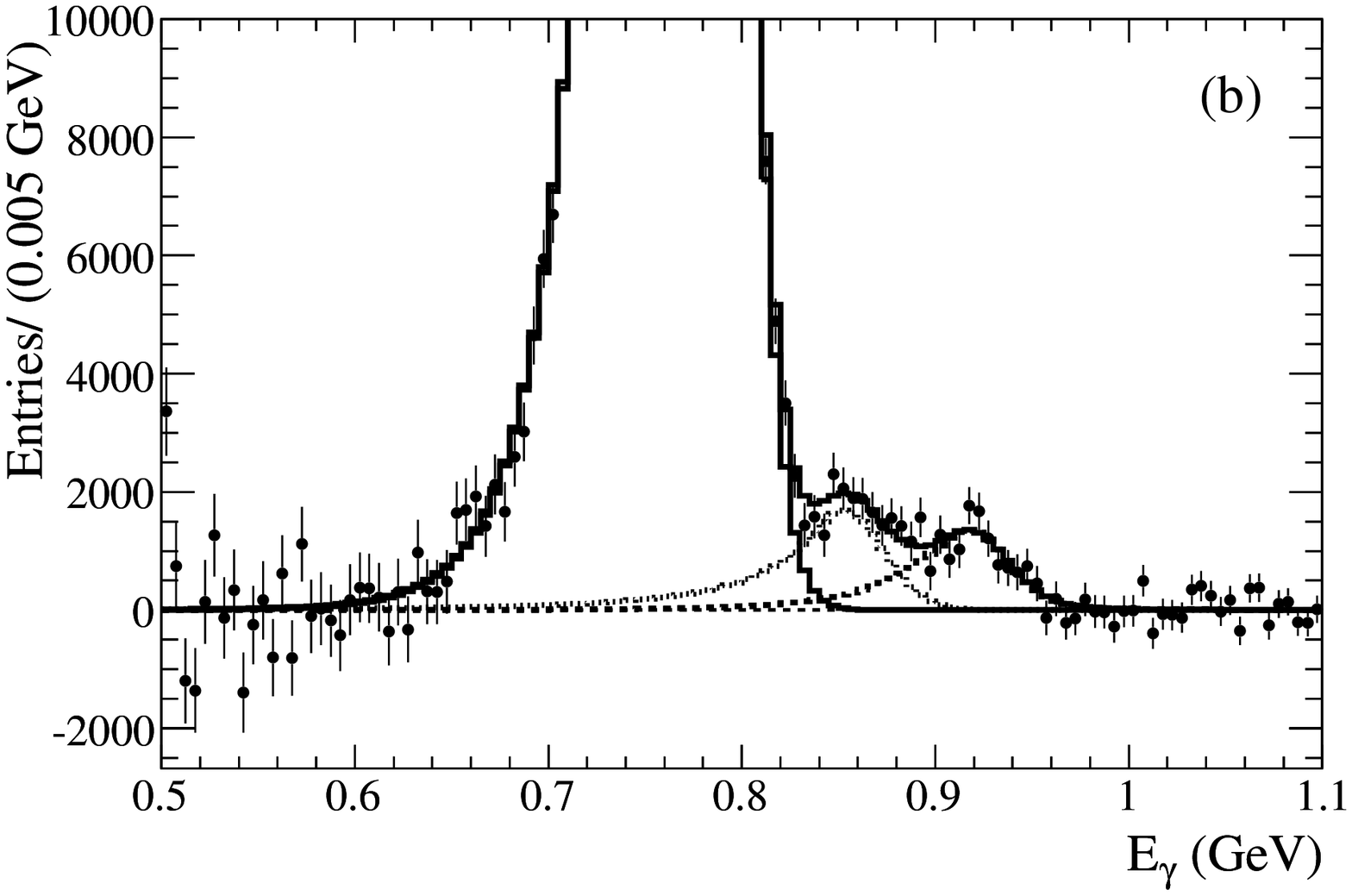}
\caption{BaBar results for the observation of $\eta_b$~\cite{ups-babar1}.  (Left) The observed inclusive photon spctrum.  (Right) The background subtracted photon spectrum.  The peaks, from left to right, are from $\chi_{bJ}$, $\Upsilon(1S)$ ISR, and $\eta_b$.}
\end{figure}


\section{Analysis of CLEO Data}


At CLEO we have an 18 times smaller data set than BaBar (5.9M compared to 109M), so several improvements in analysis procedure were needed in order to identify the $\eta_b$ signal.  The three main improvements were: performing a detailed study of the background parameterization; obtaining accurate parameterizations of the photon peak line shapes; and using the power of the thrust angle to separate signal and background to perform a joint analysis of the full data in three bins of the thrust angle.

In Fig.~2, we show the inclusive energy spectra for isolated electromagnetic showers in CLEO $\Upsilon(3S)$ and $\Upsilon(2S)$ data, after contributions from $\pi^0\to\gamma\gamma$ have been rejected.  Three features are immediately visible from these spectra.  One, the spectra are dominated by a larger, smoothly varying background from other bottomonium decays and decays from $e^+e^-\to\gamma^*\to$lighter~hadrons.  Two, the visible peaks come from the unresolved transitions $\chi_{bJ}(2P,1P)\to\gamma\Upsilon(1S)$, $J=0,1,2$.  Lastly, the peaks from $e^+e^-\to\gamma_\mathrm{ISR}\Upsilon(1S)$ and $\Upsilon(3S,2S)\to\gamma\eta_b(1S)$ (with $E_\gamma(\eta_b)\approx600$~MeV for $\Upsilon(2S)$ and $E_\gamma(\eta_b)\approx920$~MeV for $\Upsilon(3S)$) are expected to be more than an order of magnitude weaker than the $\chi_{bJ}(2P,1P)$ peaks, and they reside on the high energy tails of $\chi_{bJ}(2P,1P)$.  

\textbf{Photon Line Shapes:} Photon line shapes in crystal calorimeters, like those used by CLEO, are generally parameterized in terms of the Crystal Ball function.  This function combines a Gaussian (width $\sigma$), with a low-energy power law tail (parameters $\alpha$ and $n$).  Knowing the shape of the low energy tail is especially important for the $\chi_{bJ}$ peaks, since they are important in determing the shape of the background at the energy of the small $\eta_b$ signal.
These parameters must be determined by fitting a background-free photon peak.  In our analysis we use two independent methods.  In one method, we use the observed shape of photons from a given energy from radiative Bhabha events, and in the other we use the shape of photons from the excusive decays $\chi_{b1}(2P,1P)\to\gamma\Upsilon(1S),~\Upsilon(1S)\to l^+l^-$.  Both methods give consistent results, and once the parameters are determined, they are fixed in subsequent analyses.

\begin{figure}
\includegraphics[width=2.8in]{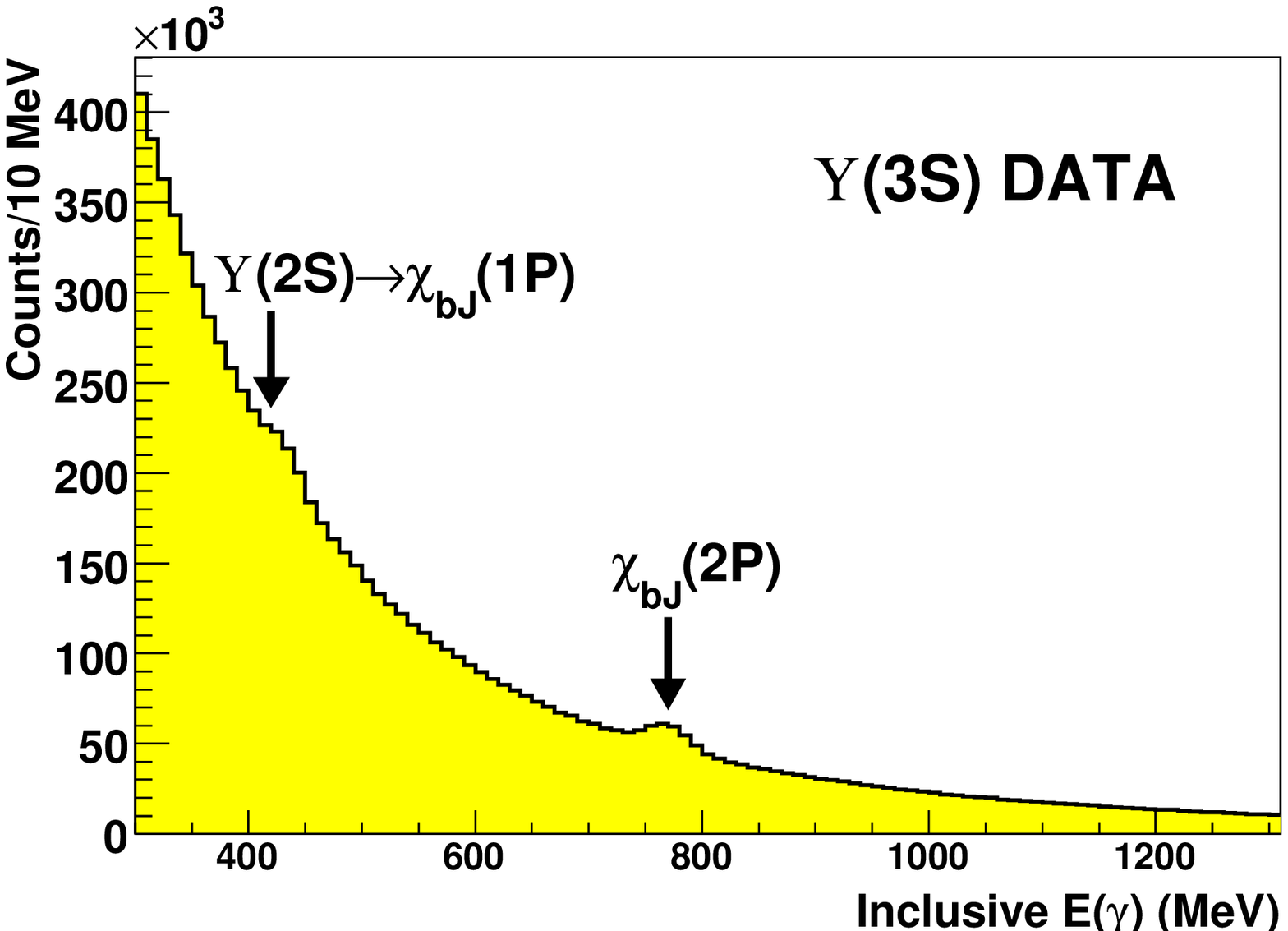}
\includegraphics[width=2.8in]{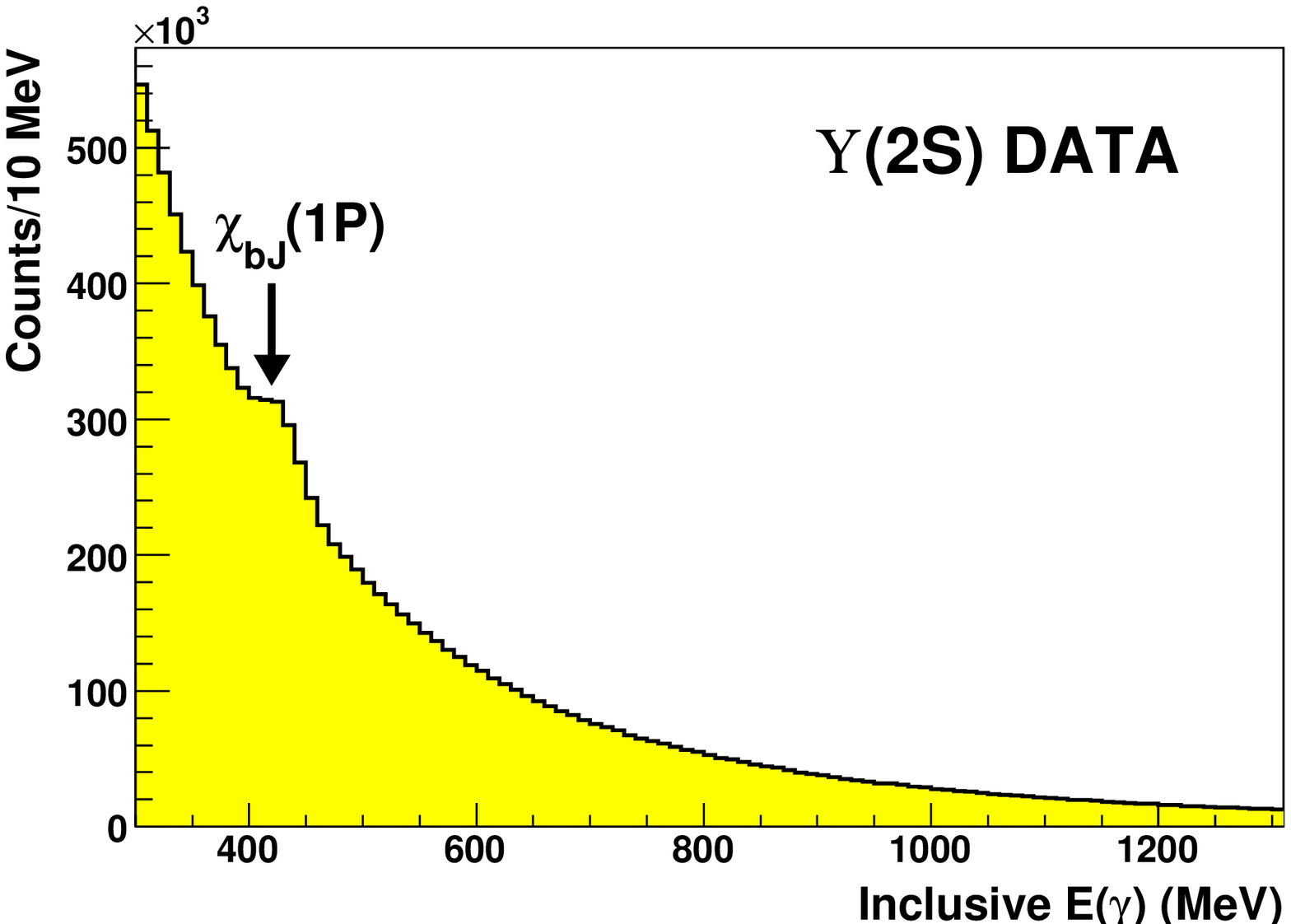}
\caption{CLEO inclusive photon spectra for (left) $\Upsilon(3S)$ decay, and (right) $\Upsilon(2S)$, illustrating the broad features described in the text.}
\end{figure}

\textbf{ISR Peak}: All the parameters of the ISR photon peak from $\Upsilon(nS)\to\gamma_\mathrm{ISR}\Upsilon(1S)$ were fixed. The yield of the ISR peak was estimated by extrapolating the observed yield in CLEO data taken at $\Upsilon(4S)$, and was then fixed to this value.

\textbf{Background}: We find that fitting the smooth background is the most crucial component in determing the results for the weak $\eta_b$ peak.  We made several hundred background fits to the data in each of three bins of $|\cos\theta_T|$, using exponential polynomials of various orders (2,3,4), in various energy regions (500--1340~MeV), and with linear and logarithmic binning of the data.  We found that many of these fit the data acceptably, and took as our final results the average of $E_\gamma(\eta_b)$, $\mathcal{B}(\Upsilon(nS)\to\gamma\eta_b)$, and significance for all the good fits (CL$>10\%$). The r.m.s. variation of these fits was then taken as a measure of the systematic uncertainty in the results from this source, $\pm1$~MeV in $E_\gamma$, $\pm10\%$ in $\mathcal{B}(\eta_b)$, and $\pm0.4\sigma$ in significance.

\textbf{Joint Analysis in Three Bins of $|\cos\theta_T|$}: As Fig.~4~(left) shows, the distribution of thrust angle ($|\cos\theta_T|$) for the background-dominated data is strongly peaked in the forward direction, but for the transition photon to the $\eta_b$, the distribution is expected to be uniform.  This happens because the transition photon is uncorrelated with the particles produced by the decay of the $\eta_b$, whereas the background photons tend to be correlated with the other particles produced in the underlying event.
We therefore define three different regions of $|\cos\theta_T|$: region~I ($|\cos\theta_T|=0-0.3$), region~II  ($|\cos\theta_T|=0.3-0.7$), and  region~III ($|\cos\theta_T|=0.7-1.0$)). Fig.~4~(right) illustrates the different levels of signal and background in the three regions.

Unlike BaBar, we do not reject events in the $|\cos\theta_T|>0.7$ region.  Instead, we simultaneously fit the spectra for each of the three regions, which lets each region contribute to the total result weighted by its individual signal-to-background.  We call this method the ``joint fit''.
We have analyzed our data by the joint fit method, and also with $|\cos\theta_T|<0.7$, for comparison with BaBar's method.  We find that the joint fit method enhances the significance of the $\eta_b$ identification by $\sim1\sigma$.

\begin{figure}

\includegraphics[width=2.9in]{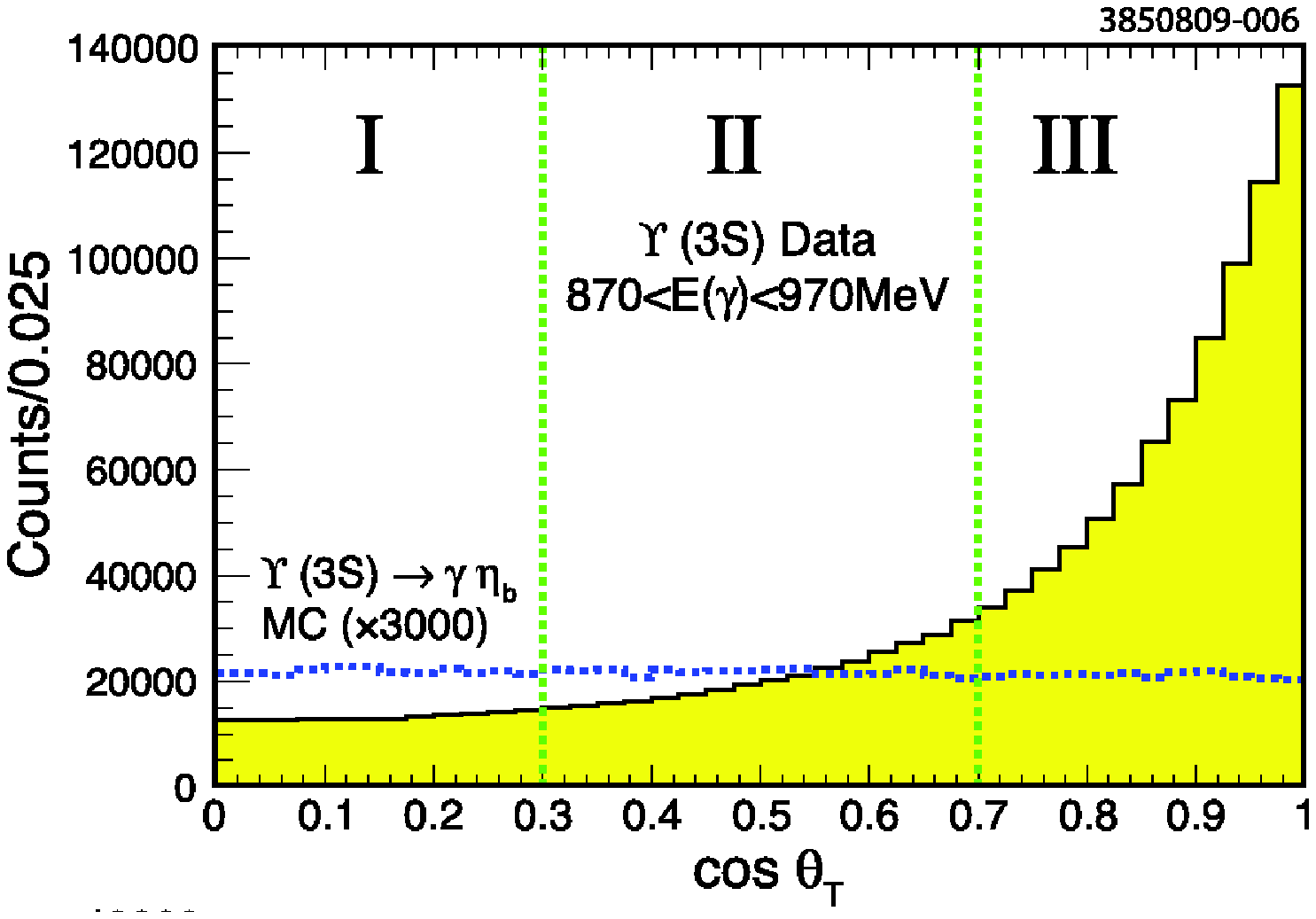}
\includegraphics[width=2.9in]{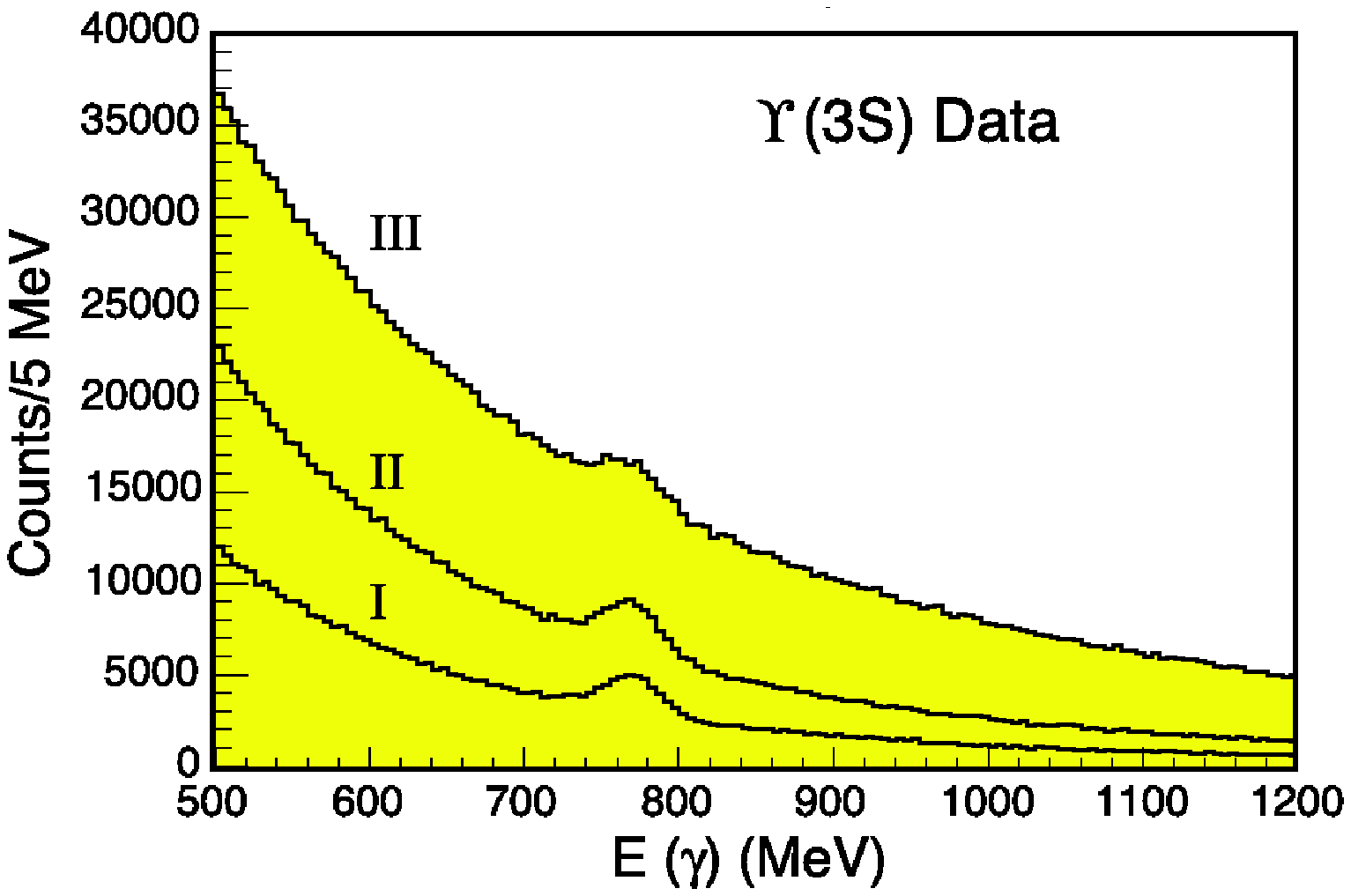}
\caption{(Left) Thrust angle $|\cos\theta_T|$ distribution for $\Upsilon(3S)$ data in the expected $\eta_b$ signal region, and the expected distribution for the $\eta_b$ signal from MC simulations.  (Right) Inclusive photon spectra for the data in the three thrust angle regions, illustrating their different signal/background ratios.}
\end{figure}

\begin{figure}
\includegraphics[width=3.5in]{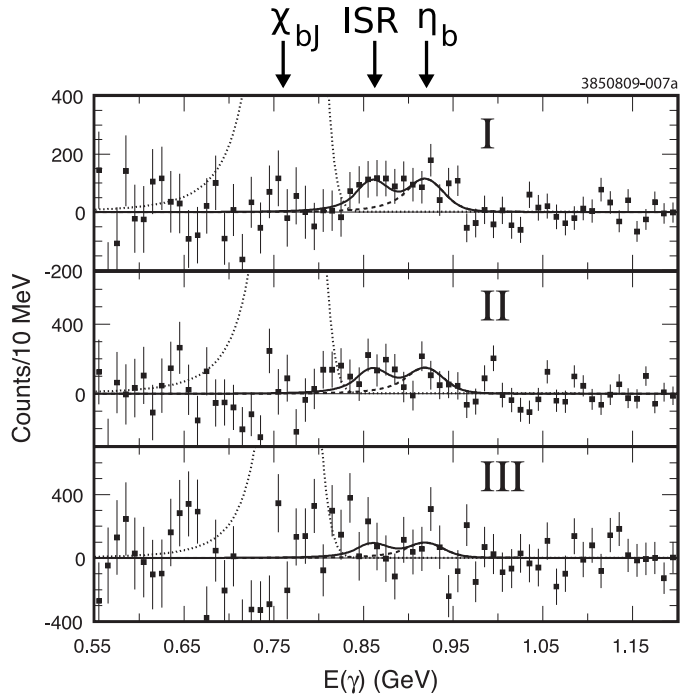}
\caption{Background subtracted spectra and representative fit result for $\Upsilon(3S)\to\gamma\eta_b(1S)$.  The $\eta_b(1S)$ peak is clearly visible in $|\cos\theta_T|$ region I and II.}
\end{figure}

\begin{figure}
\includegraphics[width=3.5in]{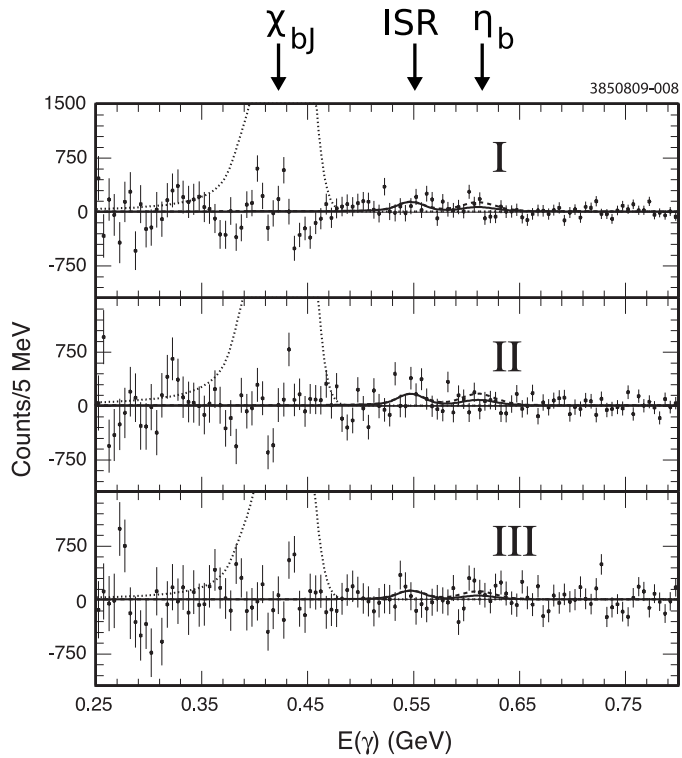}
\caption{Background subtracted spectra and representative fit result for $\Upsilon(2S)\to\gamma\eta_b(1S)$.  The dashed line corresponds to the 90\% CL upper limit given in the text.  The $\eta_b(1S)$ is not seen in any of the three regions of $|\cos\theta_T|$.}
\end{figure}

\section{Joint Fit Results}

We show a representative fit for the three bins of thrust angle in Fig.~5.
The results of the fit are:  $N(\eta_b)=2311\pm546$~counts, $E_\gamma(\eta_b)=918.6\pm6.0$(stat)~MeV, which corresponds to the hyperfine splitting of $68.5\pm6.6$~MeV, $\mathcal{B}(\Upsilon(3S)\to\gamma\eta_b)=(7.1\pm1.8(\mathrm{stat}))\times10^{-4}$, and significance $4.1\sigma$.

Although we fix $\Gamma(\eta_b)=10$~MeV, because the width of $\eta_b$ is unknown we find that the branching fraction depends on the assumed $\eta_b$ width linearly as $\mathcal{B}(\Upsilon(3S)\to\gamma\eta_b)\times10^4 = 5.8+0.13(\Gamma(\eta_b)~\mathrm{in~MeV})$.
Our results are dominated by their statistical uncertainties; the systematic uncertainties were determined conservatively and are given in Table~1.  We further note that although we quote an uncertainty in photon energy calibration of $\pm1.2$~MeV, the energies of the ISR peak and the $\chi_{bJ}(2P)$ centroid agree with their expected energies within $\pm0.3$~MeV.

\begin{table}
\begin{tabular}{lcc}
\hline 
 & \multicolumn{2}{c}{\textbf{Uncertainty in}} \\
\textbf{Source} & $\bm{E_\gamma}$ \textbf{(MeV)} & $\bm{\mathcal{B}(\Upsilon(3S)\to\gamma\eta_b)}$ \\
\hline
Background  (fn, range, binning)   & $\pm1.0$ & $\pm10\%$ \\
Photon Energy Calibration & $\pm1.2$ & --- \\
Photon Energy Resolution  & $\pm0.3$ & $\pm2\%$ \\
CB and $\chi_{bJ}(2P)$ Parameters   &    $\pm0.7$ & $\pm8\%$ \\
ISR Yield & $\pm0.4$ & $\pm3\%$ \\
Photon Reconstruction & --- & $\pm2\%$ \\
$N(\Upsilon(3S))$ & --- & $\pm2\%$ \\
MC Efficiency  & --- & $\pm7\%$ \\
\hline
\textbf{Total} &  $\pm1.8$ &  $\pm15$\% \\
\hline 
\end{tabular}
\caption{Systematic uncertainties in measurement of $\Upsilon(3S)\to\gamma\eta_b$ photon energy and branching fractions.}
\end{table}


The analysis method described for $\Upsilon(3S)$ was also used for our $\Upsilon(2S)$ data.  Because the background near the expected $\Upsilon(2S)\to\gamma\eta_b(1S)$ signal ($E_\gamma\approx610$~MeV) is approximately 6 times higher than the corresponding $\Upsilon(3S)$ (see Fig.~3) transition, no evidence for the excitation of $\eta_b$ was observed in any $|\cos\theta_T|$ bin, as illustrated in Fig.~6.  The joint analysis led to an upper limit of $\mathcal{B}(\Upsilon(2S)\to\gamma\eta_b)<8.4\times10^{-4}$, at 90\% confidence level.

We summarize our results and compare them to BaBar's results in Table~2.  Both results are in agreement.  In Table~2 we also list the theoretical predictions for hyperfine splitting and branching ratio.  The various potential model predictions vary over a wide range.  Lattice calculations for the hyperfine splitting are also given~\cite{lqcd1,lqcd2,lqcd3}, which generally agree with the experimental results.

\begin{table}
\begin{tabular}{llccc}
\hline
& &  $\Delta M_{hf}(1S)_{b\bar{b}}$, (MeV) &  $\mathcal{B}(\Upsilon(nS)\to\gamma\eta_b)\times10^4$ &  significance \\
\hline
$\Upsilon(3S)\to\gamma\eta_b$ & (CLEO)~\cite{ups-cleo-new} & $68.5\pm6.6\pm2.0$ & $7.1\pm1.8\pm1.1$ & $4\sigma$ \\
                              & (BaBar)~\cite{ups-babar1} & $71.4^{+3.1}_{-2.3}\pm2.7$ & $4.8\pm0.5\pm0.6$ & $\ge10\sigma$\\
$\Upsilon(2S)\to\gamma\eta_b$ & (CLEO)~\cite{ups-cleo-new} & --- & $<8.4$ (90\% CL) & --- \\
                              & (BaBar)~\cite{ups-babar2} & $66.1^{+4.9}_{-4.8}\pm2.0$ & $3.9\pm1.1^{+1.1}_{-0.9}$ & $3.0\sigma$ \\
\hline
\multicolumn{2}{l}{Lattice  (UKQCD+HPQCD)~\cite{lqcd1}} & $61\pm14$ \\
        & (TWQCD)~\cite{lqcd2} & $70\pm5$ \\
        & (Ehrman)~\cite{lqcd3} & $37\pm8$ \\
\multicolumn{2}{l}{pQCD (various)}    & $35-100$ & $0.05-25$ ($\Upsilon(3S)$) \\
 & & & $0.05-15$ ($\Upsilon(2S)$) \\
\hline
\end{tabular}
\caption{Summary of $\eta_b$ results from CLEO and BaBar for hyperfine splitting $\Delta M_{hf}(1S)_{b\bar{b}}$ and branching ratio, and comparison with theoretical predictions.}
\end{table}

\bibliographystyle{aipproc}   

\begin{thebibliography}{99}

\bibitem{ups-disc} S. W. Herb \textit{et al.}, \emph{Phys. Rev. Lett.}~\textbf{39}, 252~(1977).

\bibitem{etabcusb} P. Franzini \textit{et al.}, (CUSB Collaboration), \emph{Phys. Rev.}~\textbf{D 35}, 2883 (1987).

\bibitem{ups-cleo} M.~Artuso \textit{et al.} (CLEO Collaboration), \emph{Phys. Rev. Lett.}~\textbf{94}, 032001~(2005).

\bibitem{etabaleph} A. Heister \textit{et al.} (ALEPH Collaboration), \emph{Phys. Lett.}~\textbf{B 530}, 56 (2002).

\bibitem{etabdelphi} J. Abdallah \textit{et al.} (DELPHI Collaboration), \emph{Phys. Lett.}~\textbf{B 634}, 340 (2006).


\bibitem{ups-babar1} B. Aubert \textit{et al.} (BABAR Collaboration), \emph{Phys. Rev. Lett.}~\textbf{101}, 071801 (2008).

\bibitem{ups-babar2} B. Aubert \textit{et al.} (BABAR Collaboration), \emph{Phys. Rev. Lett.}~\textbf{103}, 161801 (2009).  


\bibitem{thrust} S. Brandt \textit{et al.}, \emph{Phys. Lett.} \textbf{12}, 57 (1964); E. Farhi, \emph{Phys. Rev. Lett.} \textbf{39}, 1587 (1977).

\bibitem{ups-cleo-new} G.~Bonvicini \textit{et al.} (CLEO Collaboration), submitted to \emph{Phys. Rev. Let.} [{\tt arXiv:0909.5474}].


\bibitem{lqcd1}  A. Gray \textit{et al.}, (HPQCD and UKQCD Collaborations), \emph{Phys. Rev.}~\textbf{D 72}, 094507 (2005).

\bibitem{lqcd2}  T.~Burch and C.~Ehmann,  \emph{Nucl.\ Phys.}~\textbf{A 797}, 33 (2007).

\bibitem{lqcd3}  T.~W.~Chiu, T.~H.~Hsieh, C.~H.~Huang and K.~Ogawa  (TWQCD Collaboration),  \emph{Phys.\ Lett.}~\textbf{651}, 171 (2007).


\end{thebibliography}



\end{document}